\documentstyle[11pt]{article}
\begin{document}
\vspace*{.7cm}
\begin{flushright}
\large{CERN-TH.7370/94\\
UND-HEP-94-BIG08}
\end{flushright}
\vspace{1.2cm}
\begin{center}
\LARGE{\bf OPEN QUESTIONS IN CHARM DECAYS
DESERVING AN ANSWER} \footnote{Contributed paper to appear
in the Proceedings of CHARM2000, Fermilab, June 1994}
\end{center}

\vspace{1cm}
\begin{center}\Large
I.I. Bigi\\
\vspace*{.4cm}
{\normalsize{\it TH Division, CERN, CH-1211 GENEVA 23,
SWITZERLAND \footnote{During the academic year 1993/94}} \\
and \\
{\it Dept. of Physics, Univ. of Notre Dame du Lac,
Notre Dame, IN 46556, U.S.A.\footnote{Permanent address}} \\
{\it e-mail address: IBIGI@VXCERN, BIGI@UNDHEP.HEP.ND.EDU}}
\vspace{.4cm}
\end{center}

\thispagestyle{empty}\vspace{.4cm}
\centerline{\Large \bf Abstract}
\vspace{1.0cm}
A list is given of those open questions concerning the
dynamics of charm decays where there exists a strong need
for an answer. Such a need is based on lessons to be
learnt about QCD -- either in their own right or for a better
understanding of $B$ physics -- or on searches for New Physics
with a small background from the Standard Model. The major items
on this list are: lifetimes of the $\Xi _c^{0,+}$ baryons;
semileptonic branching ratios of $D_s$, $\Lambda _c$ and
$\Xi _c$ hadrons and absolute branching ratios
for those states; radiative decays
$D\rightarrow \gamma K^*,\, \gamma \rho /\omega , \,
D_s \rightarrow \gamma \phi /\omega ,\,
D\rightarrow l^+l^-K/K^*$;
$D^0-\bar D^0$ oscillations down to a sensitivity below
$10^{-4}$ and CP asymmetries in non-leptonic $D$ decays down
to 0.1\%. Ongoing and already approved experiments will
produce important new insights, which are unlikely to
provide sufficient answers to all these questions yet. It is
discussed how a third-generation fixed-target experiment like
CHARM2000 or a $\tau$-charm factory can fill the bill.

\vspace{5cm}
\noindent CERN-TH.7370/94\\
\noindent July 1994 \\

\newpage
\large
\addtocounter{footnote}{-3}
\addtocounter{page}{-1}

One can always raise further issues about a physical
system. Yet the mere fact that some questions still
wait for an answer does not mean that there exists any real
need for obtaining those answers. My discussion will therefore
proceed in three steps: first I will list those open questions
concerning the physics of charm decays, which, in my judgement,
strongly deserve an answer; next I will try to anticipate which
of those will be answered to which degree in on-going
or already approved experiments, including those at the
asymmetric $B$ factories; in the final step I will attempt to
evaluate to which degree new initiatives, such as a new generation
fixed-target experiment -- as envisioned by CHARM2000 --
or a tau-charm factory, can make significant new contributions.

In passing I would like to note that intriguing open questions
remain also concerning charm production, such as the nature of
leading particle effects, the size of associated (i.e.
$\Lambda _c \bar D$) production and of diffractive charm
production, the specifics of charm-anticharm correlations etc.
However, I will not address these in this note.

\section{Worthy Open Questions in Charm Decays}

According to the Standard Model (SM) charm decays constitute
a decidedly dull affair: the relevant KM parameters
$V(cs)$ and $V(cd)$ are well known; for the smallness of
$|V(cb)|$ and $|V(ub)|$ constrains $V(cs)$ and $V(cd)$
very tightly through KM unitarity. Slow $D^0-\bar D^0$
oscillations, small CP asymmetries and tiny branching ratios
for rare decays are expected.

This is actually the Pessimist's perspective; the Optimist will
look at these statements and re-interpret them in a
constructive way:

\noindent $\bullet$ Because $V(cs)$ and $V(cd)$ are well-known
{\sl a priori}, one can employ charm decays to study the
workings of QCD in a novel environment under
{\sl controlled} laboratory conditions.

\noindent $\bullet$ Precisely because the SM promises us no
drama in charm decays, one can conduct searches for
$D^0-\bar D^0$ oscillations, CP violation and rare charm
decays as probes for New Physics (NP)
with an almost zero background from the SM.

\noindent $\bullet$ In addition it now appears that these
phenomena might become observable
after all at the new facilities,
even if they occur only at the level of the SM expectations.

Let me first summarize our present understanding of charm
decays:

\subsection{Lifetimes}

While most predictions of charm lifetimes have historically
turned out to be embarrassing for theory (or at least for the
authors involved), postdictions have done much better. While
this is not very surprising, it represents a non-trivial
success, if it is based on a systematic and self-consistent
treatment. Heavy Quark Expansions (HQE) provide us with such
a framework. To be sensitive to lifetime differences among
charm $mesons$, one has to go to order $1/m_c^3$. In the table
below I have juxtaposed the `Predictions' for the lifetime ratios
{}~\cite{BU,BLOKS}
with present data.

\begin{center}
\begin{tabular}{|l|l|l|}
\hline
 & QCD($1/m_c$ expansion) & Data \\
\hline
$\tau (D^+)/\tau (D^0)$ & $\sim \, 2$
(mainly due to {\em destructive} interference)
& $2.50 \pm 0.05$ \\
\hline
$\tau (D_s)/\tau (D^0)$ & $1 \pm$ few$\times 0.01$
& $1.13 \pm 0.05$ \\
\hline
$\tau (\Lambda _c)/\tau (D^0)$ & $\sim \,  0.5$ &
$0.51 \pm 0.05$ \\
\hline
\end{tabular}
\end{center}
\vspace{0.5cm}

In evaluating the theoretical entries in this table one has
to keep in mind that the theoretical
uncertainty is estimated to be
around 30\%; the observed value for $\tau (D^+)/\tau (D^0)$ is thus
reproduced within the expected errors.

Lifetimes for charm-strange baryons have been measured as well,
yet with quite unsatisfactory errors, as listed in the next table.

\begin{center}
\begin{tabular}{|l|l|l|}
\hline
 & QCD($1/m_c$ expansion)+ quark models ~\cite{BLOKS} & Data \\
\hline
$\tau (\Xi _c^+)/\tau (\Lambda _c)$ & $\sim 1.3$
& $1.68 \pm 0.5$ \\
\hline
$\tau (\Xi _c^+)/\tau (\Xi _c^0)$ & $\sim 2.8$
& $2.46 \pm 0.75$ \\
\hline
\end{tabular}
\end{center}
\vspace{0.5cm}

Considering that $m_c$ represents at best a moderately large
expansion parameter, the agreement between theoretical
expectations and present data is better than could have been
anticipated. I can identify a need for improved experimental
accuracy only in two respects: (i) Present data on the lifetimes
of $\Xi _c^{0,+}$ baryons clearly leave something to be
desired. A 10\% accuracy on $\tau (\Xi _c^{0,+})$ represents an
appropriate goal; a similar measurement of $\tau (\Omega _c)$
would be neat. Such data would provide us with valuable cross
checks of the $1/m_c$ expansion for baryon decays, yield
indirect information on terms of higher order in $1/m_c$ not
yet computed, and allow us to make numerically meaningful
extrapolations to beauty baryon lifetimes.  (ii) Measuring
the ratio $\tau (D_s)/\tau (D^0)$ with $\sim 1$\% precision
would provide us with a rather sensitive gauge for the
impact of `weak annihilation' (WA) in charm decays and for the
weight of $SU(3)_{Fl}$ breaking.

\subsection{Semileptonic Decays of Charm Hadrons}

Somewhat dated measurements read
\begin{equation}
\label{bsld+}
b_{SL}(D^+)\equiv BR(D^+\rightarrow e^+ X)=17.2 \pm 1.9\%
\end{equation}
\begin{equation}
\label{bsld0}
b_{SL}(D^0)\equiv BR(D^0\rightarrow e^+ X)=7.7 \pm 1.2\% \, .
\end{equation}
whereas a very recent CLEO analysis has yielded:
\begin{equation}
\label{CLEOII}
b_{SL}(D^0)=6.97\pm 0.18\pm 0.30 \% \, .
\end{equation}
Their ratio is consistent with the observed $D^+-D^0$ lifetime ratio.
The absolute numbers are also reproduced reasonably well in the
$1/m_c$ expansion ~\cite{BUV}.

$BR(D_s\rightarrow l X)$ has not been measured yet (only
constrained), nor have $BR(\Xi _c^{0,+}\rightarrow l X)$;
I also remain unconvinced that
$BR(\Lambda _c\rightarrow l X)$ has truly been measured. It should
be noted that while $\Gamma (D^+\rightarrow l X_s)=
\Gamma (D^0\rightarrow l X_s)$ holds, due to isospin invariance,
no symmetry argument can be invoked for
$\Gamma (\Lambda _c\rightarrow l X_s)$ vs.
$\Gamma (D^0\rightarrow l X_s)$; in the $1/m_c$ expansion one
actually finds $\Gamma _{SL}(\Lambda _c) \sim
(0.85-0.9)\times \Gamma _{SL}(D)$ through order $1/m_c^2$.

The lepton energy spectra have been measured in inclusive
$D$ decays, but not with a high degree of accuracy;
the Cabibbo-suppressed $c\rightarrow d$ transitions
have not been identified there yet.  Exclusive decays
such as $D\rightarrow l\nu K/K^*$ have been studied and
$D^0\rightarrow l\nu \pi$ have been seen.

Yet the overall data base is highly unsatisfactory and
calls for a significant improvement. The insights to
be gained from it concerning the workings of QCD would be
valuable not only in their own right, but would be a great
asset in understanding the weak decays of beauty hadrons
in general and in extracting $|V(cb)|$ and $|V(ub)|$ in
particular. To be more specific: (i) The semileptonic
widths of $D$, $D_s$, $\Lambda _c$ and preferably
$\Xi _c$ should be measured with at least 5\% accuracy.
Comparing them with each other and the corresponding
non-leptonic widths will illuminate the impact of WA.
(ii) The observed value of $\Gamma _{SL}(D)$ yields an
important calibration point for understanding the
semileptonic width of $B$ mesons as a function of $|V(cb)|$.
(iii) Analysing the lepton {\em spectra} in inclusive
semileptonic decays separately
of $D^0$, $D^+$ {\em and} $D_s$ mesons, in particular
in the endpoint region, will provide us with rather direct
information on the weight of WA and other hadronization
effects.

\subsection{Absolute Branching Ratios}

{\em Absolute} branching ratios for $D^0$ and $D^+$ decays have
been determined with 5-10\% accuracy. Nothing is known in this
respect about $\Xi_c$ and precious little about
$\Lambda _c$ decays. Reviewing events over the last two years
I feel little confident that the absolute branching ratios for
$D_s$ decays are known to better than 30\% -- if even that.

I regard this situation as truly embarrassing, since the absolute
charm branching ratios constitute an important `engineering input'
in beauty physics. The uncertainties in the charm branching ratios
are emerging as the limiting factor in determining the
branching ratios of beauty decays such as
$B\rightarrow l \nu D^{(*)}$, $B_s\rightarrow l \nu D_s^{(*)}$
and $\Lambda _b\rightarrow l \nu \Lambda _c$, with obvious
consequences for extracting a numerical value for $|V(cb)|$.
Any analysis of the charm content in $B$ decays depends on the
absolute branching ratios of charm hadrons, and any claim of a
`charm deficit' is therefore severely compromised by our
ignorance in that respect.

\subsection{Rare Decays}

An observation of $D^+,\, D_s^+\rightarrow \mu ^+\nu , \,
\tau ^+\nu$ will allow a reliable extraction of the values
for the decay constants $f_D$ and $f_{D_s}$. A battery of theoretical
estimates cluster around~\cite{SACH}
\begin{equation}
\label{fd}
f_D\sim 200 \pm 30\, \mbox{MeV}, \;
f_{D_s}\sim 200 \pm 30\, \mbox{MeV}, \;
f_{D_s}/f_D \simeq 1.15-1.2
\end{equation}
The Mark III upper bound on $D^+\rightarrow \mu ^+ \nu$ yields
$f_D \leq 290\, \mbox{MeV}$ at 90\% C.L. Recent studies
by CLEO and WA75 on $D_s\rightarrow \mu ^+ \nu$
yield $f_{D_s}=344\pm 37\pm 52\pm 42$ MeV~\cite{CLEO}
(for $BR(D_s\rightarrow \phi \pi = 3.7\%$) and
$f_{D_s}=232\pm 45\pm 20\pm 48$ MeV~\cite{WA75}, respectively.
I view these
as pilot studies, establishing in principle
that such decays can be observed and measured not only at
$D\bar D$ threshold.

The occurrence of radiative decays such as
$D\rightarrow \gamma K^*, \, \gamma \rho /\omega$ or
$D_s\rightarrow \gamma \phi , \, \gamma \rho /\omega$
{\em per se} would not be remarkable theoretically, since they
can proceed via WA coupled with photon emission off the
initial light antiquark line. Yet their observation would serve
an important ulterior motive.  For it has been suggested
\cite{ALI} that
the KM parameter $|V(td)|$ can be extracted from exclusive
radiative $B$ decays: $BR(B\rightarrow \gamma \rho /\omega)/
BR(B\rightarrow  \gamma K^*)\simeq |V(td)|^2/|V(ts)|^2$. This is
based on the assumption that both radiative transitions are
dominated by the electromagnetic penguin operator. There is
however a fly in the ointment of this interesting suggestion:
WA coupled with photon emission also generates
$B\rightarrow \gamma \rho /\omega$ transitions and this
WA contribution is independent of $|V(td)|$ and estimated to be
roughly comparable in size to the penguin contribution
\footnote{WA also contributes to $B^-\rightarrow \gamma K^{*-}$,
but that can be neglected.}! Ignoring such a contribution
would lead to the extraction of an incorrect number for
$|V(td)|$. Radiative charm decays on the other hand do not
receive any significant contributions from penguin operators,
only from WA. Measuring $BR(D\rightarrow \gamma K^*)$ and
$BR(D\rightarrow \gamma \rho /\omega)$ will provide us with
an important calibration for gauging the impact of WA on
$B\rightarrow \gamma \rho /\omega$. As a rough estimate one
expects $BR(D\rightarrow \gamma K^*) \sim
10^{-5}-10^{-4}$ and
$BR(D\rightarrow \gamma \rho /\omega) \sim
10^{-6}-10^{-5}$~\cite{OTHERS}.

There is actually a nice bonus to be found in measuring
these charm decays: New Physics can generate
$c\rightarrow u\gamma$ transitions leading to
$D\rightarrow \gamma \rho /\omega$, but not to
$D\rightarrow \gamma K^*$. Observing
\begin{equation}
\label{SUSY}
\frac{BR(D\rightarrow \gamma \rho /\omega )}
{BR(D\rightarrow \gamma K^*)}\neq
\tan ^2\theta _c
\end{equation}
would then signal the intervention of NP, of which
non-minimal SUSY is one relevant example~\cite{BGM}!

\subsection{$D^0-\bar D^0$ Oscillations}

According to the SM the rate for $D^0-\bar D^0$ oscillations
is quite slow, namely
\begin{equation}
\label{D0D0}
r_D\equiv \frac{\Gamma (D^0\rightarrow l^-X)}
{\Gamma (D^0\rightarrow l^+X)}\sim {\cal O}(10^{-4}) \, .
\end{equation}
The $D^0-\bar D^0$ transitions are driven by long-distance
dynamics within the SM; the prediction stated in
Eq.(~\ref{D0D0}) therefore suffers from considerable
numerical uncertainties. The best  available experimental
bound comes from E691:
\begin{equation}
\label{E691}
r_D\leq 3.7\times 10^{-3} \; (90\% \, C.L.)\, .
\end{equation}
There is intrinsically nothing
to prevent NP to intervene
at this level; i.e. a measurement with improved
sensitivity could reveal a positive signal. Observing a
non-vanishing value for $r_D$ between $10^{-4}$ and
$10^{-3}$ would at present not constitute irrefutable
evidence for NP, considering the uncertainties in the
SM prediction. There is some hope that those can be reduced
in the future, partly through theoretical efforts and partly
through more precise and comprehensive data on
$D^0\rightarrow K^+K^-,\, \pi ^+\pi ^-, \, K^0\bar K^0, \,
\pi ^0\pi ^0,\, \pi ^-K^+,\, K\bar K\pi , \, 3\pi , \,
K\bar K\pi \pi , \, 4\pi$ modes. For a more reliable estimate
of $\Gamma (D^0\rightarrow \bar D^0)$ can be obtained from a
dispersion relation involving the measured branching ratios for
the channels common to $D^0$ and $\bar D^0$ decays.

\subsection{CP Violation in Charm Decays}

CP asymmetries of very different forms and shapes can arise
in charm decays: they can involve $D^0-\bar D^0$
oscillations or represent direct CP violation; in the latter case
they can refer to decay widths or to final-state
distributions like $T$-odd correlations in
$D\rightarrow K\bar K \pi \pi$ modes.

\subsubsection{Direct CP Violation}

Since direct CP asymmetries require the interference of two
different weak amplitudes with different strong phases, one
has the best (and within the SM the only) chance to observe such
an effect in Cabibbo-suppressed charm decays like
$D^0\rightarrow K^+K^-, \, \pi ^+\pi^-; \;
D^+\rightarrow K^+K^-\pi ^+,\, \phi \pi ^+$. No CP asymmetry
has been observed yet, with the best bounds so far coming
from E687 and CLEO:

\begin{center}
\begin{tabular}{|l|l|l|}
\hline
 Decay mode & Measured asymmetry & 90\% C.L. limit \\
\hline
$D^0\rightarrow K^+K^-$ & $0.024\pm 0.084$ \cite{E687}&
$-11\%<A_{CP}<16\%$ \\
 &$0.071 \pm 0.065$ \cite {CLEOII}  &$-3.6\% <A_{CP}<17.8\%$ \\
\hline
$D^+\rightarrow K^-K^+\pi ^+$ & $-0.031\pm 0.068$ \cite{E687}
& $-14\%<A_{CP}<8.1\%$ \\
\hline
$D^+\rightarrow \bar K^{*0}K^+$ & $-0.12\pm 0.13$ \cite{E687}&
$-33\%<A_{CP}<9.4\%$ \\
\hline
$D^+\rightarrow \phi \pi ^+$ & $0.066\pm 0.086$ \cite{E687}&
$-7.5\%<A_{CP}<21\%$ \\
\hline
$D^0\rightarrow K_S\phi$ &$-0.005\pm 0.067$ \cite{CLEOII}
&$-11.5\%<A_{CP}<10.5\%$ \\
\hline
$D^0\rightarrow K_S\pi ^0$ &$-0.011\pm 0.030$ \cite{CLEOII}
&$-6\%<A_{CP}<3.8\%$ \\
\hline
\end{tabular}
\end{center}
\vspace{0.5cm}

The requirement to encounter strong final-state interactions
does not pose any problem in principle, since charm decays
proceed in the resonance region below 2 GeV; yet at the
same time it introduces an element of considerable numerical
uncertainty into the predictions. A rough estimate suggests
that within the SM direct CP asymmetries could be as `large'
as ${\cal O}(10^{-3})$~\cite{BTAUCHARM,GOLDEN}.
Fitting a set of quark diagrams to
describe a host of non-leptonic two-body modes of $D$ mesons
leads to quite a similar conclusion~\cite{BUCELLA}.
It is not inconceivable
that NP could enhance these asymmetries somewhat, say to the
1\% level.

Larger effects could surface in the Dalitz plots for
$D\rightarrow K\bar K \pi , \, 3\pi$ or in $T$-odd correlations,
like for example
$\langle \vec p_{\pi ^{\pm}}\cdot (\vec p_{K^+}\times \vec p_{K^-}
)\rangle$ in $D^{\pm}\rightarrow K^+K^-\pi ^{\pm}\pi ^0$.

\subsubsection{CP Asymmetries involving $D^0-\bar D^0$
Oscillations}

In the presence of $D^0-\bar D^0$ oscillations and for a channel
$f$ common to $D^0$ and $\bar D^0$ decays, the required
interference can occur between the amplitudes for
$D^0\rightarrow f$ and $\bar D^0\rightarrow f$. Examples
for such final states are $f=K^+K^-,\, \pi ^+\pi ^-, \,
K_s\pi ^0, \, K_S\omega , \, K_S\eta$. Ignoring the
possibility of direct CP violation one writes down:
\[
\Gamma (D^0\rightarrow f; t)=
e^{-\Gamma _Dt}|T(D^0\rightarrow f)|^2
(1-Im \frac{q}{p}\bar \rho _f\sin \Delta m_Dt) \]
\begin{equation}
\label{CPOSC}
\Gamma (\bar D^0\rightarrow f;t)=
e^{-\Gamma _Dt}|T(\bar D^0\rightarrow f)|^2
(1+Im \frac{q}{p}\bar \rho _f\sin \Delta m_Dt)
\end{equation}
with $\bar \rho _f=T(\bar D^0\rightarrow f)/
T(D^0\rightarrow f)$, denoting the ratio of decay amplitudes
and $q/p$ reflecting $D^0-\bar D^0$ oscillations. Three
observations should be noted here \cite{BS}:

\noindent (i) While this CP asymmetry becomes unobservable
for $\Delta m_D=0$, it actually is proportional to
$\Delta m_D/\Gamma _D$ for small values of $\Delta m_D$. The
quantity $r_D$, introduced in eq.(~\ref{D0D0}), on the other hand
is given by $\frac{1}{2}(\Delta m_D/\Gamma _D)^2$.
(For simplicity I ignore $\Delta \Gamma _D$ effects
although, within the SM, one expects very roughly
$\Delta \Gamma \sim {\cal O}(\Delta m_D)$.) Thus the
experimental bound on $r_D$ translates into
$\Delta m_D\leq 0.09\cdot \Gamma _D$ and the CP asymmetry
\begin{equation}
\label{ASYM}
A_{CP}^f\equiv \frac{\Gamma (\bar D^0\rightarrow f; t)
-\Gamma (D^0\rightarrow f; t)}
{\Gamma (\bar D^0\rightarrow f; t)
+\Gamma (D^0\rightarrow f; t)}
\simeq \frac{\Delta m_D}{\Gamma _D}\frac{t}{\tau _D}
Im \frac{q}{p}\bar \rho _f
\end{equation}
could still reach values of several per cent!

\noindent (ii) No such luck arises in the SM: for reasons
that are quite specific to it, one finds
$\Delta m_D\sim {\cal O}(0.01)\Gamma _D$ and
$Im (q/p)\bar \rho \sim {\cal O}(10^{-3})$; i.e. the
size predicted by the SM for these kinds of asymmetry is
presumably too small to be observable.

\noindent (iii) Accordingly one should vigorously search
for CP asymmetries involving $D^0-\bar D^0$ oscillations:
their dependance on the (proper) time of decay provides a
striking experimental signature; observing them --
as defined in eq.(~\ref{ASYM}) -- with a size of $10^{-3}$
or above constitutes a clear sign for the intervention of
NP.

Hence we arrive at the following benchmarks concerning future
studies of CP violation: one should aim for achieving a
$10^{-3}$ sensitivity for CP asymmetries involving
$D^0-\bar D^0$ oscillations as well as for direct CP
violation. Observation of an effect unequivocally
signals the presence of NP in the former case, but not
necessarily in the latter.

\section{Answers Expected To Be Obtained by Existing or
Approved Experiments}

Over the next four years I expect important new data to come from
experiments at FNAL, CERN, Beijing and
Cornell. In five years from now the asymmetric $B$ factories at
KEK and SLAC will start to contribute. I anticipate the most
significant new information in the following areas:

\noindent (i) A more precise determination of $\tau (D_s)$,
and the
first fully quantitative measurement of $\tau (\Xi _c^{+,0})$.

\noindent (ii) The first measurement of
$BR(D_s\rightarrow l+X)$ and studies of the {\em inclusive}
lepton spectrum in semileptonic $D_s$ decays; the first
{\em direct} determination of $BR(D_s\rightarrow \phi \pi )$.

\noindent (iii) Extracting the {\em absolute} values of
$BR(D\rightarrow K\pi , \, K\pi \pi )$ to better than 5\%.

\noindent (iv) Possibly a measurement of {\em absolute}
$\Lambda _c$ branching ratios via a
$\Sigma _c\rightarrow \Lambda _c \pi $ tag.

\noindent (v) The first quantitative extraction of $f_D$ and
$f_{D_s}$ from $D,\, D_s \rightarrow \mu \nu$.

\noindent (vi) Mapping out the doubly-Cabibbo-suppressed
$D$ and $D_s$ decays.

\noindent (vii) A rather comprehensive analysis of
Cabibbo-favoured and once-Cabibbo-suppressed $D$, $D_s$ and
possibly $\Lambda _c$ decays.

\noindent (viii) Detailed studies of exclusive semileptonic
decays $D\rightarrow l \nu K/K^*/\pi /\rho ,\,
D_s\rightarrow l \nu \eta /\phi / K/K^*$ and
$\Lambda _c\rightarrow l \nu \Lambda /\Sigma$, with the
dependance of the form factors on the momentum transfers
measured rather than assumed.

\noindent (ix) A probe of $D^0-\bar D^0$ oscillations down to
$r_D\sim 10^{-4}$ and CP asymmetries down to a few per cent.

All these anticipated data will certainly deepen our
understanding of the hadrodynamics driving charm decays:

\noindent (a) Applying a comprehensive BSW-type analysis of the
two-body modes of $D$ and $D_s$ mesons (and preferably of
$\Lambda _c$ baryons as well) {\em separately} to
Cabibbo-allowed, once- and twice-suppressed decays will undoubtedly
reveal clear deviations from the predictions based on
factorization, presumably with a definite pattern. It will also help
us to arrive at better estimates of $\Delta m_D|_{SM}$, and it
will sharpen our understanding of where we can expect the largest
direct CP asymmetries, and what size they can reach within the SM.

\noindent (b) It will be immensely instructive to compare detailed
data on exclusive semileptonic $D$, $D_s$ and $\Lambda _c$
decays with predictions obtained in particular through
simulations of QCD on the lattice.

\noindent (c) The improved accuracy in the measurements of
$\tau (D_s)$ and $\tau (\Xi _c^{0,+})$ will provide us with a
handle to arrive at a quantitative understanding of charm
lifetimes and at the same time with a gauge from which to
extrapolate to $\bar \tau (B_s)$, $\tau (\Lambda _b)$
and $\tau (\Xi _b)$.

\noindent (d) Observing $D^0-\bar D^0$ oscillations and/or
CP violation would represent a major discovery; its ramifications
would of course depend on the numerical size of the effect.

Yet, despite all this progress, major tasks will remain
unaddressed or at least unfinished: (i) A $\sim 5\%$ measurement
of $\tau (\Omega _c)$ would be quite helpful, although this is not
the major item among the unfulfilled tasks. (ii) I find it
doubtful that the {\em absolute} branching ratios for
$D_s$, $\Lambda _c$ or $\Xi _c$ decays will have been determined
within even 10\%. (iii) Likewise, $f_D$ and $f_{D_s}$ will not
have been measured to better than 20\% or so. (iv) Nothing useful
will be known about the radiative decays
$D\rightarrow \gamma K^*/\rho /\omega , \,
D_s \rightarrow \gamma \phi /\rho /\omega$. (v) The accuracy
will still be unsatisfactory, with which the total semileptonic
widths will be known for $D$, $D_s$ and $\Lambda _c$, let alone
for $\Xi _c$; likewise for the {\em inclusive} lepton spectra.

At first sight, this list might appear like a rather pitiful
collection of small morsels having fallen off the main table. In
particular, I have already implied that I expect all
two-body channels of $D$ and $D_s$ mesons to have been measured
with sufficient accuracy and detail, i.e. including modes with
one or two neutrals. Yet I would like to state quite emphatically
that the above list represents very major unresolved problems
using the criteria given in the introduction:

\noindent $\bullet$ Weak decays of charm
hadrons constitute a microscope
to study the strong interaction effects crucial for a full
understanding and thus exploitation of beauty decays.

\noindent $\bullet$ Charm decays provide a rather clean lab to search
for manifestations of NP in rare $D$ decays, $D^0-\bar D^0$
oscillations and CP violation.

These two aspects will not have been treated with the `ultimate'
sensitivity. I therefore conclude: in all likelihood there
will remain a strong and identifiable need for another major new
initiative for studies of charm decays to understand hadronization
effects down to the level of the QCD `noise' and to probe for NP
down to the SM `noise' -- or to better understand a signal
that has emerged!

\section{New Initiatives for the Next Millenium}

I will attempt to evaluate the potential of two
complementary facilities to provide the `final' answers
in the physics of charm decays, namely CHARM2000 on the one hand
and a $\tau$-charm factory on the other.

\subsection{CHARM2000}

A next-generation experiment based on fixed-target production of
charm will be able to do a superb job in measuring the
{\em relative} branching ratios of a host of exclusive
non-leptonic channels in $D$, $D_s$, $\Lambda _c$ and $\Xi _c$
accurately. I am however not convinced at all that our
understanding of charm decays would improve in proportion, since
I am sceptical that the theoretical `noise', i.e. the
irreducible uncertainties, will drop to the per cent level.
I should add one caveat, though: I could see a
meaningful progress emanate from CHARM2000 measurements of
(quasi-)two-body modes if previous experiments -- contrary to my
expectations stated above -- had failed to measure channels
containing two neutrals in the final state with decent accuracy.

In my opinion there are then five main challenges against which
the significance and the merits of CHARM2000 can be judged:

\noindent (1) The lifetimes of $\Xi _c$ and
preferably also of $\Omega _c$ baryons should be measured
with an accuracy of at least 5\%.

\noindent (2) The decay constants $f_D$ and $f_{D_s}$
should be extracted from $D,\, D_s\rightarrow \mu \nu$
to within 10\%.

\noindent (3) CHARM2000 would again have the statistical
muscle to observe the radiative decays
$D\rightarrow \gamma K^*/\rho /\omega ,\,
D_s\rightarrow \gamma \phi /\rho /\omega$
(and also
$D\rightarrow l^+l^-K/ K^*/\rho /\omega$, etc.) at the transition
rate expected for them. The question is whether backgrounds
like $D\rightarrow \pi ^0K^*\rightarrow \gamma [\gamma ]K^*$
can be controlled.

\noindent (4) Can {\em absolute} branching ratios be
determined to within $\sim 1-2\%$ for $D$,
within $\sim 5\%$ for $D_s$ and within $\sim 10\%$ for
$\Lambda _c$ decays? The strong decay
$D^*\rightarrow D\pi$ can be used for calibrating
the $D$ branching ratios; for the other charm hadrons
new calibration methods have to be pioneered, like
$\Sigma _c\rightarrow \Lambda _c\pi$.

\noindent (5) Can $D^0-\bar D^0$ oscillations be probed down
to $r_d\sim 10^{-5}$, which almost certainly should reveal a
positive signal? Even more crucially, can systematics be
controlled to such a degree that a comprehensive search
for CP asymmetries involving $D^0-\bar D^0$ oscillations
and direct CP violation can be undertaken with a
sensitivity of $10^{-3}$ or even smaller?

There is another aspect to be briefly mentioned,
not -- in all fairness -- as a formal challenge, but
rather as a potential bonus of quite significant weight:
(i) Can the inclusive semileptonic widths of the different
charm hadrons be measured with, say, 5\% accuracy? \break
(ii) Can
the lepton energy spectra in inclusive semileptonic charm
hadron decays be measured with an accuracy that allows the
extraction of the value of $|V(cd)|$ from the endpoint region?

\subsection{$\tau$-Charm Factory}

The capabilities of a $\tau$-charm factory are quite
complementary to those of CHARM2000. Clearly charm lifetimes
cannot be measured directly. What can be done -- and can be
done quite well -- is to measure semileptonic branching ratios.
For the isospin partner $D^+$ and $D^0$ one has:
$\tau (D^+)/\tau (D^0)\simeq BR_{SL}(D^+)/BR_{SL}(D^0)$. Yet
such a relation does not hold in general for all hadrons;
in particular one expects
$\tau (\Lambda _c)/\tau (D^0)\neq BR_{SL}(\Lambda _c)/
BR_{SL}(D^0)$.
Using tagged decays one can determine the {\em absolute}
branching ratios of the various charm hadrons in a clean way.
The lepton energy {\em spectra} in inclusive semileptonic
decays both of mesons and of baryons can be studied quite
well. Employing beam energy constraints should allow one to
measure radiative charm decays such as
$D\rightarrow \gamma K^*/\rho /\omega$ rather reliably. Relying
on quantum mechanical EPR-like correlations, one can probe for
$D^0-\bar D^0$ oscillations, CP asymmetries involving them
and direct CP violation \cite{BTAUCHARM}.

While all this appears feasible in principle, I see
two challenges on a practical level:

\noindent (1) Can $r_D$ be probed down to values
$\sim 10^{-5}$? Even more importantly, can one acquire the
sensitivity to search for $\sim 10^{-3}$ CP asymmetries?

\noindent (2) The clean environment at a $\tau$-charm factory
has its price: very little charm physics can be done
`parasitically'; i.e., $D$, $D_s$, $\Lambda _c$ and $\Xi _c$
decays have to be studied at different beam energies corresponding
to $D\bar D$, $D^*\bar D/D\bar D^*$, $D_s\bar D_s$,
$\Lambda _c\bar \Lambda _c$ and $\Xi _c\bar \Xi _c$ final states.
The required statistics has then to be accumulated in the
rather limited amount of time available at each beam energy --
and these beam energies have to span, merely for charm physics,
the region from the $D\bar D$ threshold up to at least the
$\Lambda _c$ and very preferably the $\Xi _c$ threshold!

\section{Summary}

There is a strong and well-defined need for another new
generation of charm decay experiments, like CHARM2000 and
a $\tau$-charm factory. Very specific challenges
can be formulated, which these projects have to overcome.
Since their approaches, strengths and drawbacks are quite
complementary, it would be wonderful if both could be
realized.

\vspace{1cm}

\noindent {\bf Acknowledgements}

\noindent I have learnt a lot from many
discussions with my collaborators
B. Blok, M. Shifman, N. Uraltsev and A. Vainshtein; I have also
benefitted from exchanges with T. Mannel, J. Cumalat,
D. Kaplan, A. Nguyen and P. Sheldon.
This work was supported by the National Science Foundation under
grant number PHY 92-13313.

\end{document}